# TUNABLE ANTENNA-COUPLED INTERSUBBAND TERAHERTZ (TACIT) MIXERS: THE QUANTUM LIMIT WITHOUT THE QUANTUM LIQUID

M. S. Sherwin, C. Cates, B. Serapiglia, Y. Dora, and J. B. Williams

Physics Department and Center for Terahertz Science and Technology, UCSB, Santa Barbara, CA 93106

K. Maranowski and A. C. Gossard

Materials Department, UCSB, Santa Barbara, CA 93106

W. R. McGrath

Center for Space Microelectronics Technology, Jet Propulsion Laboratory, California Institute of Technology, Pasadena, CA 91109

## ABSTRACT

At present, nearly quantum-limited heterodyne receivers for Terahertz (THz) frequencies require cooling to temperatures below 4K. We are working to build semiconductor-based "tunable antenna-coupled intersubband terahertz" (TACIT) heterodyne mixers for 1-5 THz. We predict that they can achieve single-sideband noise temperatures of a few hundred K with intermediate frequency (IF) bandwidth >10 GHz, while operating at lattice temperatures >20 K and requiring local oscillator (LO) power ~ 1 µW.

## INTRODUCTION

It is widely recognized that the technology for submillimeter wavelengths, or Terahertz frequencies, is far behind the technology at lower and higher frequencies. In the last few years, breakthroughs in performance have been achieved with superconducting hot-electron bolometers (HEBs)[1-4]. These have demonstrated noise temperatures within an order of magnitude of the quantum limit at frequencies > 1 THz for LO powers less than 1 µW, with multi-GHz IF bandwidths. However, superconducting HEBs must be cooled to temperatures below 4K, requiring liquid Helium or rather complex and bulky mechanical coolers. In this paper, we describe a novel semiconductor-based THz mixer, the tunable antenna-coupled intersubband terahertz (TACIT) mixer. Modeling predicts that TACIT mixers are capable of noise temperatures within a factor of ~5 of the quantum limit for frequencies >1 THz with intermediate frequency (IF) bandwidths exceeding 10 GHz. TACIT mixers will require less than 1 µW of local oscillator (LO) power, achievable with compact solid state sources, and operate at temperatures T≥20K, achievable with existing lightweight mechanical coolers. Thus, successful development of TACIT mixers would enable, for the first time, relatively small and inexpensive, long-duration space-science missions which incorporate nearly quantum-limited heterodyne receivers for THz frequencies.

*Contact information for M. S. Sherwin: Email: sherwin@physics.ucsb.edu, phone (805)893-3774*

## QUALITATIVE DISCUSSION

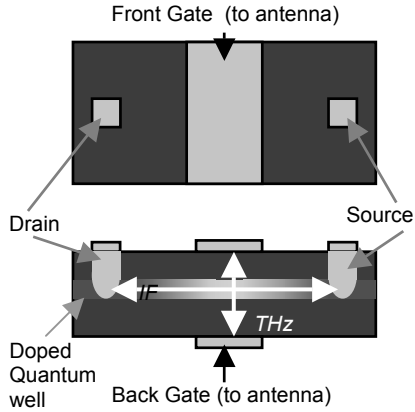

*Fig. 1: Schematic diagram of active region of TACIT mixer.*

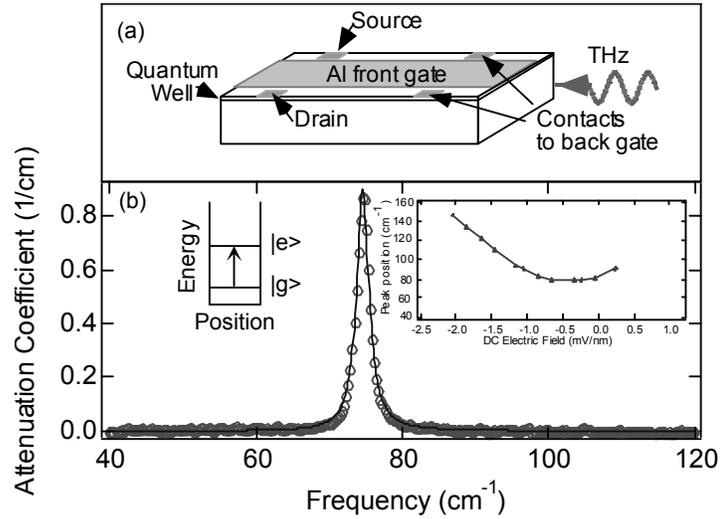

*Fig. 2(a) Schematic diagram of sample processed for characterization. (b) Attenuation vs. frequency. The resonance is associated with the |g> to |e> intersubband transition of a 40 nm square quantum well (left inset). The right inset shows tuning of the intersubband absorption peak with a dc electric field.*

Fig. 1 shows a schematic diagram of the active region of a TACIT mixer in top and side view, respectively. THz radiation impinging on a coplanar antenna (not shown) induces an electric field oscillating at THz frequencies between a front and back gate. DC voltages may also be applied to these gates. When the THz frequency is resonant with an intersubband transition, it is efficiently absorbed by a 2-D electron gas (see below), causing it to warm up. This heating of the electron gas results in a change in its resistance, which is sensed via a current applied between a source and a drain. Two-terminal millimeter-wave mixers have been demonstrated based on heating a 2-D electron gas, but their high conversion loss has made them non-competitive for operation at THz frequencies.[5,6]

Fig. 2 a shows a schematic diagram of a GaAs wafer which has been processed for characterization of its intersubband absorption. An Al front gate, and ohmic contacts to a buried doped quantum well and buried back-gate have been deposited. The quantum well in this case is a 40 nm wide GaAs layer, with barriers made of $Al_{0.3}Ga_{0.7}As$. The Si dopants which provide electrons to the quantum well are located remotely, in the $Al_{0.3}Ga_{0.7}As$ barriers. This reduces scattering, giving high mobility and narrow intersubband absorption linewidths[7]. THz radiation is incident to the edge of the sample, with electric field polarized in the growth direction (perpendicular to the plane of the quantum well). The symbols in Fig. 2b show a strong peak in the absorption, centered at 75 cm$^{-1}$ (2.3 THz) with a full-width-at half-max (FWHM) of about 2 cm$^{-1}$ (0.06 THz), derived from the Lorentzian fit (solid line). This peak corresponds to the transition between ground and first excited "subbands" of the quantum well, as shown schematically in the left inset to Fig. 2b. The TACIT mixer will absorb only photons at this intersubband transition frequency. The right inset shows the tunability of the center frequency with a dc electric field created by varying the voltage between the front and back gates, allowing one to choose the frequency at which the TACIT mixer detects. Experiments on similar samples show that the time $\tau_e$ for the population to relax from the excited to the ground state varies from roughly 1 ns at temperatures less than 10K to 10 ps at 50 K[8].

# MODEL

## Absorption

The antenna or transmission line which couples THz radiation into the active region of the TACIT mixer will typically have an impedance > 50 Ohms. We show here that the active region of the TACIT mixer can be designed (and tuned in-situ with gate voltages!) to nearly match this impedance, resulting in near unit coupling efficiency between antenna and active region. This impedance calculation has not been presented before, and will be expanded upon in a future publication.

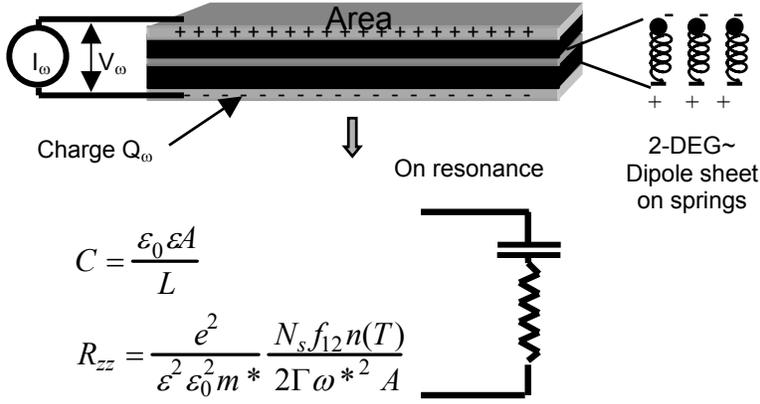

*Fig. 3: Illustration of the model used for computing the impedance of the active region on resonance.*

To compute the impedance of the active region, we consider an oscillating current $I_\omega$ flowing from the antennas onto the front and back gates (which can be thought of as plates of a capacitor), and compute the induced oscillating voltage drop $V_\omega$ between the plates. An oscillating charge $Q_\omega$ on the gates induces an electric field $E_\omega = Q_\omega/(A\varepsilon\varepsilon_0)$ in between them, where $A$ is the area of the gates, $\varepsilon$ is the dielectric constant (13 in GaAs), and $\varepsilon_0$ is the permittivity of free space. We treat the 2-D electron gas as a 2-D system in which a 2-D polarization density $P_{2D}$ can be induced by the electric field inside the capacitor:

$$P_{2D} = \chi_{2D}(\omega) E_\omega. \tag{1}$$

An induced 2-D polarization will reduce the voltage drop across the capacitor. If the distance between the gates is $L$, the voltage drop between them will be given by

$$V_\omega = E_\omega L - P_{2D}/\varepsilon\varepsilon_0. \tag{2}$$

Assuming that the current $I_\omega = i\omega Q_\omega$, and making appropriate substitutions, the impedance presented by the active region is given by

$$Z(\omega) = \frac{V_\omega}{I_\omega} = \frac{1}{i\omega\varepsilon\varepsilon_0 A}\left(L - \frac{\chi_{2D}(\omega)}{\varepsilon\varepsilon_0}\right) \tag{3}$$

The theory of the 2-D susceptibility $\chi_{2D}(\omega)$ associated with intersubband transitions in quantum wells has been worked out in the late 1970s.[9,10] This theory takes into account the dynamic screening of the external oscillating electric field $E_\omega$ inside the quantum well. In the absence of such dynamic screening, the absorption frequency would simply be given by $E_{12}$, the energy required to promote a single electron from the ground subband to the excited subband at constant in-plane momentum. However, the intersubband absorption resonance occurs at a higher energy $\hbar\omega*$ because one is exciting a collective mode of the system (all the electrons must oscillate together). In the so-called Hartree approximation, the shift $\hbar\omega* - E_{12}$ is often called the *depolarization shift*. Remarkably, all of the effects of dynamic screening and electron-electron interaction on $\chi_{2D}(\omega)$ can be folded into two parameters, the frequency $\omega*$ and the oscillator strength $f_{12} = 2m* z_{12}^2 E_{12}/\hbar^2 \leq 1$, where $m* = m_0/15$ is the effective mass of an electron in GaAs. In the Hartree approximation, the matrix element of the position operator $z_{12}$ is computed using wave functions for subband 1 and 2 which satisfy self-consistently the Schrödinger and Poisson equations. The expression for $\chi_{2D}(\omega)$ is

$$\chi_{2D}(\omega) = \frac{N_s e^2 f_{12} n(T)}{m*} \frac{1}{\omega*^2 - \omega^2 + i2\omega\Gamma}. \tag{4}$$

Here, $e$ is the electron charge, $\omega$ is the angular frequency of the oscillating electric field, and $2\pi\Gamma$ is the HWHM of the intersubband absorption. The normalized population difference $n(T)$ accounts for the thermal saturation of the intersubband transition, which occurs when the temperature is sufficiently high to significantly populate subband 2. The quantity $n(T)=(N_1(T)-N_2(T))/N_S$ where $N_1(T)$ ($N_2(T)$) is the population of subband 1 (subband 2), and $N_S$ is the total sheet density. If one treated the intersubband absorption as arising from a sheet of harmonic oscillators with charge $e$ (as shown in Fig. 3), each oscillating at frequency $\omega*$ with HWHM=$2\pi\Gamma$ and ignored depolarization effects, one would come up with an expression identical to (4), but with $f_{12}=n(T)=1$.

For $\omega=\omega*$, the equivalent circuit for the active region is a capacitor in series with a resistor, as shown in Fig. 3, with impedance

$$Z(\omega*) = \frac{V_{\omega*}}{I_{\omega*}} = \frac{L}{i\omega*\varepsilon\varepsilon_0 A} + \frac{N_S e^2 f_{12} n(T)}{\varepsilon^2 \varepsilon_0^2 A m^* \omega^{*2} 2\Gamma} = \frac{1}{i\omega*C} + R_{zz} \qquad (5)$$

where $C$ is the capacitance associated with the active region. Note that, in this device, resistance $R_{zz}$ to THz currents flowing perpendicular to the plane of the quantum well is very different than resistance to IF currents which flow in the plane of the quantum well.

The coupling efficiency from a transmission line with impedance $R_{zz}$ into a load with matched impedance is given by (5) is $4R_{zz}^2/(4R_{zz}^2+X^2)$, where X is the capacitive reactance $1/(i\omega C)$. Table 1 shows that coupling efficiencies close to 90% can be achieved with reasonable device parameters.

## Responsivity and noise temperature

Once THz radiation is coupled into the active region, there are a number of ways in which it could alter the resistance between source and drain, which we call $R_{xx}$. We consider here a simple model in which the energy deposited by THz radiation is quickly thermalized in the active region, causing the entire electron gas to warm up to a temperature higher than that of the lattice. Since the source-drain resistance varies with electron temperature, the absorbed THz radiation can be detected.

Following Karasik,[11] and assuming perfect coupling of RF into and IF out of the active region, we compute the mixer noise temperature using the expression

$$T_N = \frac{NEP^2}{\alpha 2 P_{LO} k_B} \qquad (6)$$

Here $P_{LO}$ is the incident local oscillator power, $k_B$ is Boltzmann's constant, $\alpha$ is the coupling efficiency, and NEP is the noise-equivalent power of the device operating as an incoherent detector (i. e., with no local oscillator), in W/Hz$^{1/2}$. Considering Johnson noise and noise associated with temperature fluctuations of the active region, the NEP is given by

$$NEP^2 = NEP^2_{JOHNSON} + NEP^2_{TF}$$
$$= \frac{4k_B T R_{xx}}{\alpha \rho^2} + 2k_B T_e^2 G(T_e) \alpha^{-1} \qquad (7)$$

where $4k_B TR$ is the Johnson and shot noise voltage power, in V$^2$/Hz, $\rho$ is the responsivity, in V/W, and $G(T_e)$ is the thermal conductivity. The second term is the noise associated with fluctuations in the temperature of the active region. Shot noise is insignificant in the metallic transport we consider here. Following Karasik,[11] we assume that the active region is at a much higher electron temperature than the heat bath (leads and lattice), and hence ignore the temperature fluctuations in those reservoirs. This results in the temperature fluctuation noise being $2k_B T_e^2 G(T_e)$ rather than the usual $4k_B T_e^2 G(T_e)$.

The responsivity of the hot-electron bolometer can be simply written as

$$\rho = \frac{dV}{dP} = I\frac{dR}{dP} = \frac{V_0}{R}\frac{dR}{dT_e}\left(\frac{dP}{dT_e}\right)^{-1} = \frac{V_0 \gamma}{G(T_e)} \qquad (8)$$

where $I$ is the bias current, $V_0$ is the voltage drop in response to this current, $T_e$ is the electron temperature, $P$ is the power, and $\gamma \equiv \frac{1}{R_{xx}}\frac{dR_{xx}}{dT_e}$.

We now estimate the power input to the bolometer from both dc bias and local oscillator required to raise the electron temperature to $T_e$.

$$P_{dc} + \alpha P_{LO} = \frac{V_0^2}{R_{xx}} + \alpha P_{LO} = \int_{T_L}^{T_e} G(T)dT \approx G(T_e)T_e \qquad (9)$$

for $G(T_e)T_e \gg G(T_L)T_L$, which will be the case in our system. Here, $P_{dc}$ is the dc power which is dissipated *in the active region*. In general, the lowest noise temperature arises when the $P_{dc}=\alpha P_{LO}$, which we assume from now on. The thermal conductivity will have contributions from phonons and diffusion. The phonon contribution $G(T)=C_V/\tau=Nk_B/\tau=n_S A k_B/\tau$. Here, $C_V$ is the heat capacity of the 2-D electron gas, for which we have assumed the classical value of $Nk_B$. The intersubband relaxation time $\tau \sim 10$ ps at 50K, and decreases with increasing temperature. The IF bandwidth associated with this time is $1/2\pi\tau=16$ GHz. Diffusion cooling can produce even greater bandwidths.[6]

The classical contributions to the noise temperature of the TACIT mixer are then given by the remarkably simple expression

$$T_N = \left(\frac{8}{\gamma^2 T_e} + 2T_e\right)\alpha^{-2} \qquad (10)$$

where the first term is associated with Johnson noise, and the second with thermal fluctuations.

The temperature-dependence of the resistance of a 2-D electron gas has been investigated. A summary plot is given in Davies, p. 360.[12] In all but the very best samples, the mobility is relatively independent of temperature below 10K, where it is dominated by elastic scattering from static impurity or roughness potentials, and decreases with increasing temperature. In the very best samples, the mobility is limited by electron-phonon scattering. In such samples, near 50K, electron-LA phonon scattering dominates, and the mobility is proportional to $T^{-1.5}$.[13] With such a temperature dependence, $|\gamma(50K)|=1.5/T=0.03$. Such a value would result in a classical contribution to the noise temperature of 330K.

Table 1 shows an example of device parameters and predicted specifications for a TACIT mixer based on a 40 nm wide GaAs square quantum well with $Al_{0.3}Ga_{0.7}As$ barriers which was extensively characterized by Williams et. al.[7] The device would run at an electron temperature of 50K, with lattice temperature assumed much smaller (for example, 20K). The reduced population difference $n(T)=0.57$ was computed assuming Fermi statistics governed the occupations of subbands whose energies were determined self-consistently in the Hartree approximation.

The combination of small local oscillator power, high operating temperature, wide bandwidth, and low noise would make such a device extremely attractive. Fabrication of a prototype TACIT mixer designed to operate at 1.6 THz is under way.

*Table 1: Parameters and predicted specifications for TACIT mixer based on a 40 nm GaAs/AlGaAs quantum well with zero external electric field.*

| Area $A$ | 5 μm² | Energy relaxation time $\tau$ | 10 ps |
|---|---|---|---|
| Gate separation $L$ | 0.3 μm | IF bandwidth | 16 GHz |
| Sheet density $N_S$ | $1.3 \times 10^{15}$ m⁻² | $1/R\, dR/dT=\gamma$ | 0.03 K⁻¹ |
| Eff. osc. strength $f_{12}n(T)$ | 0.51 | Electron temperature $T_e$ | 50K |
| Linewidth $\Gamma$ | 66 GHz | SSB Noise temperature (Eq. 10) | 330K |
| RF impedance (Eq. 5) | 49 + 36i Ohms | Intersubband absorption freq. $\omega*/2\pi$ | 2.3 THz |
| Coupling efficiency | 88% | Local oscillator power | 0.5 μW |

## ACKNOWLEDGEMENT

M.S.S would like to thank Prof. Jonas Zmuidzinas for extremely helpful discussions. These resulted in a formula for $R_{zz}$ which is corrected with respect to that published previously.[14, 15]

This work was supported by NASA contract NAG5-10299.